\documentstyle[12pt]{article}

\newcommand{\lsim}{\raisebox{-0.13cm}{~\shortstack{$<$ \\[-0.07cm] $\sim$}}~}
\newcommand{\gsim}{\raisebox{-0.13cm}{~\shortstack{$>$ \\[-0.07cm] $\sim$}}~}
\newcommand{\ra}{\rightarrow}
\newcommand{\ee}{e^+e^-}

\newcommand{\nn}{\noindent}

\newcommand{\beq}{\begin{eqnarray}}
\newcommand{\eeq}{\end{eqnarray}}

\newcommand{\tb}{\tan\beta}

\begin{document}
\topskip 2cm 

\begin{titlepage}

\begin{flushright}
KA--TP--17--96\\
May 1996 \\
\end{flushright}

\vspace*{.3cm}

\begin{center}
{\large\bf PHYSICS PROSPECTS AT FUTURE e$^+$e$^-$ COLLIDERS$^*$}\\
\vspace{2.5cm}
{\large Abdelhak Djouadi}\\
\vspace{.5cm}
{\sl Institute f\"ur Theoretische Physik, Universit\"at Karlsruhe,}\\
\vspace*{.1cm}
{\sl D--76128 Karlsruhe, Germany}\\
\vspace{1.5cm}

\begin{abstract}
\nn I discuss the physics prospects at future high--energy $e^+ e^-$
linear colliders. After summarizing some aspects of these future
machines, I will review a few topics in the Standard Model which can be
studied at these colliders: top quark, $W/Z$ vector bosons and Higgs
particles physics. A brief discussion on the exploration of extensions
of the Standard Model such as Supersymmetry and Grand Unified Theories
at $\ee$ colliders will also be given. 
\end{abstract}

\end{center}

\vspace*{1.5cm}

\nn $^*$ Invited talk given at {\it Les Rencontres de Physique de la 
Vall\'ee d'Aoste}, La Thuile Aosta Valley, March 3--9 1996

\end{titlepage}
\setcounter{page}{2}

\section{Introduction}

\subsection{The Physics Motivations} 

The $\ee$ colliders which have operated in the past twenty years, from
SPEAR to LEP1 and SLC, have been very successful in searching for the
fundamental constituents of matter and in exploring their interactions.
In particular, the recent high--precision studies of the properties of
the $Z$ boson at LEP1 and SLC, have laid a very solid base for the
Standard Model (SM) of the electroweak and strong interactions. However,
within the SM, many problems remain to be solved and despite of the
spectacular agreement with the present experimental data, extensions of
the model are anticipated. 

$\bullet$ The top quark, recently discovered at the Tevatron, has a mass
of $\sim 180$ GeV. This high value renders a thorough analysis of the
properties of this particle mandatory: since the mass is close to the
Fermi scale, the top quark might hold important clues to the electroweak
symmetry breaking mechanism. 

$\bullet$ The strength and the form of the electroweak $W$ and $Z$ 
boson self--couplings are determined by the non--abelian gauge symmetry 
of the SM. However, deviations from these values are expected in more 
general scenarios. These couplings can be measured at existing colliders 
only with large errors, and the precision must be improved by one or two 
orders of magnitude to unambiguously establish the nature of these particles. 

$\bullet$ The Higgs particle, and the electroweak symmetry breaking
mechanism, are the most important ingredients of the SM. The search for
the Higgs particle will be a major task and it is very important that no
way of escape is left for this particle. Once discovered, its
fundamental properties, spin--parity quantum numbers and couplings to
the other particles, must be precisely determined. 

$\bullet$ Supersymmetry (SUSY) is the most attractive extension of the
SM. It not only stabilizes the huge hierarchy between the weak and the
GUT scales against radiative corrections, but also if SUSY is broken at
a sufficiently high scale, it allows us to understand the origin of the
hierarchy in terms of radiative gauge symmetry breaking. Moreover, SUSY
models offer a natural solution to the dark matter problem and allow for
a consistent unification of the all known gauge couplings. Many new
particles are predicted in these theories; the search for these states
and the study of their properties will be one of the major goals of
present and future colliders. 

$\bullet$ Many other possibilities for extensions of the SM can be anticipated. 
For instance, the SM group SU(3)$\times$SU(2)$\times$U(1) is expected to 
be embedded in a unifying group at a high--energy scale. In this case, 
extra gauge bosons with masses not too far from the Fermi scale might 
exist. In these unifying groups, many new matter particles are also predicted 
and some of them might be relatively light. These new gauge and matter 
particles have also to be searched for at present and futur colliders. 

While strongly interacting particles, as well as new gauge bosons, can 
be easily searched for at hadron colliders, $\ee$ colliders provide 
unique opportunities to search for non--colored particles such as Higgs
bosons, SUSY partners of leptons and electroweak/Higgs bosons and, 
due to a clean experimental environment, to make detailed 
studies of their properties. Even for strongly interacting
(s)particles and new gauge bosons, the $\ee$ colliders can play an
important role in pining down their properties. 

In this talk, I will discuss the potential of future high--energy $e^+
e^-$ linear colliders to address these issues, after summarizing some
aspects of the $\ee$ reaction and the future machines. The talk is based
on the work done at the various workshops on $\ee$ colliders [1--6] which 
took place in the last years to which I refer for original references. 

\subsection{The e$^+$e$^-$ Collision} 

The $\ee$ collision is a very simple reaction, with a well defined
initial state and rather simple topologies in the final state. It
has a favorable signal to background ratio, leading to a very clean
experimental environment which allows to easily search for new phenomena
and to perform very high--precision studies as has been shown at
PEP/PETRA and more recently at LEP/SLC. 

The physical processes are mainly mediated by $s$--channel photon and
$Z$ boson exchanges with cross sections which scale like $1/s$, and
$t$--channel gauge boson or electron/neutrino exchange, with cross
sections which rise like $\log(s)$. The $s$--channel exchange is the
most interesting process: it is democratic in the sense that it gives
approximately the same rates for weakly and strongly interacting
(matter) particles and for the production of known and new particles. 

However the rates are low at high energies, and one needs to increase
the luminosity to compensate for the $1/s$ drop of the interesting cross
sections. At $\sqrt{s} \sim 1$ TeV, one needs a luminosity of ${\cal L}
\sim 10^{34}$ cm$^{-2}$s$^{-1}$, or equivalently an integrated luminosity 
of $\int
{\cal L} \sim 100$ fb$^{-1}$ per year, to produce 10.000 muon pairs as
at PEP and PETRA. Such a luminosity is necessary to allow for thorough
data analyses, including cuts on the event samples and allowing for
acceptance losses in the detectors. At higher energies, the luminosity
should be scaled up as $s$ to generate the same number of events. 

\subsection{Aspects of the Colliders} 

Various reference designs of future high--energy $\ee$ colliders are
being studied at DESY, SLAC, KEK, Novosibirsk and CERN. In the following, 
let me briefly list a few points about these future colliders [1--3]: 

$\bullet$ Because of synchrotron radiation which rises as the fourth
power of the c.m. energy in
circular machines, $\ee$ colliders beyond LEP2 must be linear machines,
a type of accelerator which has been pioneered by the SLC. 

$\bullet$ These machines will be probably realized in two steps. I a 
first phase, they will operate in the energy range between 300 and 500 
GeV with a yearly luminosity of ${\cal L} \sim 50$ fb$^{-1}$, and in  
a second phase with a center of mass energy up to $\sqrt{s} \sim 2$ 
TeV and an integrated luminosity of ${\cal L} \sim 200$ fb$^{-1}$.
The c.m. energy of the colliders can be adjusted in order to make
detailed studies: for instance to scan the $t\bar{t}$ threshold or to 
maximize the cross section for Higgs boson production. 

$\bullet$ The requirement of a high--luminosity is achieved by squeezing
the electrons and positrons into bunches of very small transverse size,
leading to beamstrahlung which results into beam energy loss and the smearing
of the initially sharp $\ee$ c.m. energy. Since the precise knowledge of
the initial state is very important for precision studies, beamstrahlung
should be reduced to a very low level, as it is already the case in narrow 
beam designs. 
 
$\bullet$ The longitudinal polarization of the electron (and possibly
positron) beam should be easy to obtain as it has already been shown at
the SLC. The longitudinal polarization might be very important when it
comes to make very precise measurements of the properties of the top
quark and the $W/Z$ bosons for instance. 

$\bullet$ Last but not least, the $\ee$ collider can run in three
additional modes. First, one just needs to replace the positron bunches
by electron bunches to have an $e^- e^-$ collider. Then, by illuminating
the electron and positron bunches by laser photons, one can convert the
$\ee$ collider into an $e \gamma$ or $\gamma \gamma$ collider with a 
comparable total c.m. energy and luminosity as the initial $\ee$ 
collider. These modes will be very useful to address problems such
as the Higgs--photons coupling and the properties of the $W$ bosons. 

I the rest of this talk, I will nevertheless mainly focus on the $\ee$ option 
of the collider and discuss only the physics potential at the first phase with 
a center of mass energy around $\sqrt{s} \sim 500$ GeV. Occasionally, I 
will comment on the benefits of raising the energy or changing the option. 

\section{SM Physics}

\subsection{The Top Quark and QCD}

The first --and most obvious-- thing to do at a 500 GeV $\ee$ collider
is the study of the properties of the top quark. Since the top quark is
very massive, $m_t \sim 180$ GeV, it might play a very important role in
the understanding of the mechanism of electroweak symmetry breaking and
the properties of this particle should be studied in detail. 

In $\ee$ collisions, top quarks are mainly produced through $s$--channel
$\gamma$ and $Z$ boson exchange, $\ee \ra \gamma, Z \ra t\bar{t}$. The
maximum cross section is reached at about 50 GeV above threshold, i.e. at
$\sqrt{s} \sim 400$ GeV, Fig.1. A rate of 40.000 top quark pairs can be
produced with a luminosity of ${\cal L}\simeq 50$ fb$^{-1}$, a sample
comparable to the number of top quark which can be used for physics at
the LHC, after filtering out the background. The top quarks will mainly
decay into $b$ quarks and $W$ bosons, $t \ra bW^+$, and in the clean 
environment of $\ee$ colliders, the top quarks can be easily isolated. 

The very sharp rise of the $t\bar{t}$ production cross near threshold,
due to the well--know Coulomb singularity, would allow a very
high--precision measurement of the top quark mass. Conservative
estimates of the theoretical and experimental uncertainties lead to an
error of about 200 MeV on $m_t$, more than an order of magnitude
smaller than what is expected to be the case at the LHC. This precision
will allow the first stringent constraint on the mass of the Higgs boson
if this particle has not been found. 

Because of its large mass, the top quark will decay before it forms
$t\bar{t}$ bound states, and the $t$ and $\bar{t}$ will separate at
distances less than $10^{-2}$ fm. However, the $\ee \ra t \bar{t}$ cross
section near the production threshold is still sensitive to the
quark--antiquark binding potential. Therefore, the threshold region will
be the best place to study QCD: the binding is strong, but it is
completely determined by perturbation theory. From the QCD potential,
the strong coupling constant $\alpha_s$ could be determined with an
error of less than 0.4\%, and this would be the most precise individual
measurement of this coupling. 

The measurement of $m_t$ and $\alpha_s$ from the excitation curve will
be however correlated. To decorrelate the two measurements, one can
determine the momentum distribution of the top quark, the average of
which is proportional to the product of the two quantities, $<p> \sim
m_t \alpha_s$. The top quark width [as well as $m_t$ and $\alpha_s$] can
be determined from the forward--backward asymmetry. Indeed, the
interference between S and P--wave contributions near threshold is
rather strong, leading to a FB asymmetry of about 10\%. For fixed $m_t$
and $\alpha_s$, this allows the determination of $\Gamma_t$ with a
precision of less than 20\%. Note that the QCD coupling constant can
also be determined independently by measuring the annihilation cross
section into jets, allowing for a stringent test of asymptotic freedom
down to distances of ${\cal O}( 10^{-3}$) fm. 

Because of the short lifetime of the top quark, its polarization is not
lost during the decay making possible a general helicity analysis of
the angular distribution of the process $\ee \ra t\bar{t} \ra b\bar{b}
WW \ra b\bar{b}$ + 4--fermions. This will allow the determination of the
electroweak $t\bar{t}$ production and the weak ($tb$) decay currents. The
anomalous magnetic moment, the electric dipole moment [a sign of CP
violation in the $t$ sector] and the chirality of the ($tb$) current
[left--handed in the SM] can be determined at a high level. This allows
to constrain many extensions of the SM, such as multi--Higgs doublet
models, models of dynamical symmetry breaking and compositeness, which
might manifest themselves first in the top sector due the large value 
of $m_t$. 

Finally, it might well be that the top quark does not decay only into
$bW$ final states. Due to the large value of $m_t-m_b$, there
could be enough phase space for exotic decay modes to occur. For
instance, in supersymmetric extensions of the SM, the top quark
could decay into a $b$ quark and a charged Higgs boson, $t \ra b 
H^+$, or into the lightest neutralino $\chi_1^0$ [which is the lightest 
supersymmetric particle] and the scalar top squark $\tilde{t}$ [which in many
scenarios is lighter than the $t$ and all the other squarks],
$t \ra \tilde{t} \chi_1^0$. Even if these exotic decay modes are allowed by
phase space, they are expected to be rare, with branching ratios of the 
order of a few  percent. This rates would be frequent enough, 
however, to be analysed thoroughly at $\ee$ colliders. 

\subsection{W and Z Bosons} 

The gauge symmetries of the SM completely determine the form and the
strength of the self-interactions of the electroweak bosons. Deviations
of the triple $WW \gamma, WW Z$ and quartic $W^4, W^2Z^2, W^2 \gamma^2$
couplings, as well as new couplings $Z^3, Z^2 \gamma^2$ in addition to
the usual SM couplings, could be expected in more general scenarios. For
instance in composite models and models where the $W/Z$ bosons are
generated dynamically or interact strongly with each other, these
couplings can be altered. 

The couplings $WW \gamma$ and $WW Z$ are in general described by seven
parameters each. Assuming C, P and T invariance in the bosonic sector,
the number of parameters is reduced to three: $g_{\gamma, Z}$,
$\kappa_{\gamma, Z}$ and $\lambda_{\gamma,Z}$. $g_{\gamma, Z}$ are the
electric and $Z$ charges of the $W$ bosons and the parameters
$\kappa_{\gamma, Z}$ and $\lambda_{\gamma,Z}$ are related to the
corresponding magnetic dipole moments and electric quadripole moments of
the $W$ bosons. Symmetries of the underlying dynamical interactions may
further constrain these couplings. The gauge symmetries of the SM,
demand that $g_\gamma=1, g_Z= -\tan^2 \theta_W, \kappa_{\gamma, Z}=1$
and $\lambda_{\gamma,Z}=0$ at tree--level. 

The magnetic dipole and electric quadripole moments can be measured
directly in the production of electroweak bosons at hadron and $\ee$
colliders. Modifications of the self--couplings vertices destroy the SM
unitarity cancellations between interfering boson and fermion exchange
amplitudes for longitudinal vector boson production. As a result, small
deviations of the moments from their SM values are magnified by 
a coefficient $\beta_W^2 s /M_W^2$ where $\beta_W$ is the velocity
of the $W$ bosons. Because the constraints also scale with the
accumulated luminosity, a 500 GeV $\ee$ collider with a luminosity
of ${\cal L}\sim 50$ fb$^{-1}$ will have a sensitivity which is two
orders of magnitude larger than LEP2. 

The main reaction to be exploited at high--energy $\ee$ colliders is $W$
pair production $\ee \ra W^+W^-$. A large number, ${\cal O} (10^5)$, of
$W$ pairs will be produced in a clean environment for the energy and
luminosity specified above, so that a very high reconstruction
efficiency can be achieved. Additional and complementary information can
be obtained from $WW$ fusion to $Z$ bosons, $\ee \ra \nu \bar{\nu}WW
\ra \nu \bar{\nu}Z$, and in particular if laser induced $\gamma$ beams
are available, from $\gamma e^- \ra \nu W^-$ and $\gamma \gamma \ra
W^+ W^-$. These processes are separately sensitive to $\gamma$
and $Z$ couplings, and can help to disentangle $\kappa_\gamma,
\lambda_\gamma$ from $\kappa_Z, \lambda_Z$. 

From the combined analysis of the $W$ pair production cross section and
the angular distribution of the $W$ decay products [which allow to
separate the various $W$ helicity components], very tight constraints on
the anomalous couplings can be obtained. If all $\kappa$'s and
$\lambda$'s are allowed to vary freely, they can be probed to an
accuracy of $|\kappa_\gamma -1|, \, |\kappa_Z-1, \, |\lambda_\gamma|, \,
|\lambda_Z| \lsim 0.01$. More stringent bounds can be obtained if the
anomalous couplings are constrained by additional symmetry
requirements. To disentangle $\gamma$ from $Z$ couplings in this
reaction, longitudinal polarization is very important. 

The analysis of $W \gamma$ final states at the Tevatron leads to bounds
of ${\cal O}(1)$ on the anomalous photonic couplings. At the LHC,
bounds of order $|\lambda_\gamma|  \lsim 0.02$ and $|\kappa_\gamma-1|
\lsim 0.1$ are expected in this process, while information on $\kappa_Z$
and $\lambda_Z$ is more difficult to extract. 
The bounds on $|\kappa_\gamma-1|$ and $|\lambda_\gamma|$ that can be
reached at an $\ee$ collider operating at $\sqrt{s}=500$ GeV with a
luminosity of 10 fb$^{-1}$ are compared with the expected values 
at LEP2 [with a very high luminosity $\int {\cal L}=1.3$ fb$^{-1}$]
and LHC [which has the same reach as the late SSC in the
high--luminosity option $\int {\cal L}=100$ fb$^{-1}$]; Fig.~2. $\ee$
colliders are more powerful by not only probing $\kappa_\gamma$ and
$\lambda_\gamma$ to a higher accuracy, but also in providing stringent
constraints on $\kappa_Z$ and $\gamma_Z$. Note that the constraints
become more stringent at higher energies or higher luminosities. 

Various quartic vector boson couplings can also be probed at $\ee$
colliders: the $W^4, W^2Z^2$ and $Z^4$ couplings in the
reaction $\ee \ra WW Z$ and $ZZZ$, and the $W^2 \gamma^2$ and
$\gamma^2 Z^2$ couplings in the reaction $\ee \ra WW \gamma$
and $\ee \ra ZZ\gamma$. The new physics effective energy scale 
which can be probed at a 500 GeV collider is of the order of a TeV.

Finally, $WW$ scattering cannot be studied at energies around 500 GeV. 
This is a very important process to study if light Higgs particles do 
not exist and $W$ bosons become strongly interacting at high energies. 
Raising the c.m. energy of the collider to $\sqrt{s} \sim 1.5$ to 2 TeV, 
allows to study in detail this strongly interacting scenario.

\subsection{The Higgs Particle} 

\nn {\bf a) The Higgs in the SM}
\vspace*{3mm}

The most important mission of a future high--energy collider will be the
search for scalar Higgs particles and the exploration of the electroweak
symmetry breaking mechanism. In the SM, one doublet of complex scalar 
fields is needed to spontaneously break the SU(2)$\times$U(1) symmetry.
Among the four initial degrees of freedom, there Goldstones will be 
absorbed by the $W^\pm$ and $Z$ bosons to get their masses, and the 
remaining degree of freedom will correspond to a physical scalar 
particle, the Higgs boson. 

Since the couplings of the Higgs boson to fermions and gauge bosons are
proportional to the masses of these particles, the only unknown
parameter in the SM is the Higgs boson mass, $M_H$. It is a free
parameter and the only thing we know about it is that it should be
larger than $\sim 65$ GeV, from the negative searches at LEP1, and that
it is probably smaller  than 1 TeV, from fits of the high--precision
LEP1 data. [For $M_H \gsim 1$ TeV, the gauge bosons would interact
strongly to insure unitarity in the scattering of the electroweak gauge
bosons; the residual final state interaction can be studied at c.m.
energies beyond 1 TeV]. 

However, there is a theoretical argument which indicates that the Higgs
boson might be light, $M_H \lsim 200$ GeV. Indeed, the quartic Higgs
coupling is proportional to $M_H^2$ and since the scalar sector of the
SM is not an asymptotically free theory, the coupling will grow with the
energy until it reaches the Landau pole, where the theory does not make
sense anymore. If the cut--off $\Lambda$ where  new phenomena should
occur is of ${\cal O}(1$ TeV), the Higgs mass should be smaller than
$\sim 700$ GeV [as verified by simulations on the lattice]; Fig.3. But if one
wants to extend the SM up the GUT scale $\Lambda \sim 10^{16}$ GeV [a
prerequisite for the perturbative renormalization of the electroweak
mixing angle from the GUT symmetry value 3/8 down to the experimentally
observed value at low energies], $M_H$ is restricted to values smaller
than $\sim 200$ GeV. In addition, radiative corrections due to top quark
loops could drive the Higgs self--coupling to negative values, therefore
destabilizing the vacuum. The stability and the triviality bounds,
constrain the Higgs boson mass to lie in the range $ 100$ GeV $ \lsim
M_H \lsim$ 200 GeV; Fig.~3. 

A Higgs bosons with $M_H \gsim 100$ GeV, cannot be produced at LEP2 at 
a c.m. energy of $\sqrt{s}=192$ GeV and the discovery of the Higgs boson
will have to wait for the LHC. At hadron colliders, Higgs particles will
be copiously produced; however, because of the huge QCD background, one
has to rely on very rare Higgs decay channels to see the signal. In the
low mass range, $M_H \lsim 140$ GeV, the Higgs will mainly decay into
$b\bar{b}$ final states [and to a lesser extent to $\tau^+ \tau^-$,
$c\bar{c}$ and gluon pairs] and one has to use the clean $\gamma \gamma$
decay channel which has a branching ratio of ${\cal O}(10^{-3})$. For
larger masses, where the Higgs decays mainly into $W$ pairs, the $ZZ$
decays [where one of the $Z$ bosons can be virtual] become important,
but one needs to make the $Z$ bosons decays into charged leptons which
brings again the branching to the level of a few times $10^{-3}$. 

This would nevertheless be sufficient to discover the Higgs particles
at LHC after a few years of running with a rather high luminosity and
with dedicated detectors [to resolve for instance the narrow $\gamma
\gamma$ peak for $M_H \lsim 140$ GeV]. But it will be very difficult to
make a detailed study of the properties of the Higgs bosons and to
test unambiguously the mechanism of electroweak symmetry breaking. 
In contrast, an $\ee$ collider with a c.m. energy of 500 GeV will
be the ideal machine to produce Higgs particles with masses between
100 and 200 GeV and to study in detail their fundamental properties. 

\vspace*{3mm}
\nn {\bf b) Higgs Production in e$^+$e$^-$ Collisions}
\vspace*{3mm}

The main production mechanism of Higgs particles in $\ee$ collisions are 
the Higgs--strahlung process, $\ee \ra (Z^*) \ra HZ$ [with a cross section
which scales as $1/s$ and therefore dominates at low energies] and the $WW$ 
fusion mechanism, $\ee \ra \nu \bar{\nu} (W^* W^*) \ra \nu \bar{\nu}H$ [with 
a cross section rising like $\log(s/M_H^2)$ and which dominates at high
energies]. At $\sqrt{s} \sim 500$ GeV, the two processes have approximately 
the same cross sections for the interesting range 100 GeV $\lsim M_H 
\lsim$ 200 GeV; Fig.~4. With an in integrated luminosity $\int {\cal L}
\sim 50$ fb$^{-1}$, approximately 2000 events per year can be collected
in each channel; a sample which is more than enough to discover the
Higgs boson and to study it in detail. The $ZZ$ fusion mechanism, $\ee
\ra \ee (Z^* Z^*) \ra \ee H$, and the associated production with top
quarks, $\ee \ra t\bar{t}H$ have much smaller cross sections. But these
processes will be very useful when it comes to study the Higgs
properties as will be discussed later. 

In the Higgs--strahlung process, $\ee \ra HZ$, the recoiling $Z$ boson
[which can be tagged through its clean $\mu^+ \mu^-$ decay mode e.g.] is
mono--energetic and the Higgs mass can be derived from the energy of the
$Z$ boson if the initial $e^+$ and $e^-$ beam energies are sharp
[beamstrahlung, which smears out the c.m. energy should be thus suppressed as
strongly as possible, and this is already the case for machine designs
such as TESLA]. Therefore, it will be easy to separate the signal from
the backgrounds. For low Higgs masses, $M_H \lsim 140$ GeV, the
main background will be $\ee \ra ZZ$. The cross section is large, but it
can be reduced by cutting out the forward and backward
directions [the process is
mediated by $t$--channel electron exchange] and by selecting $b\bar{b}$
final states by means of $\mu$--vertex detectors [while the Higgs decays
almost exclusively into $b\bar{b}$ jets in this mass range, BR$(Z \ra
b\bar{b}$) is small, $\sim 15\%$]. The background from single $Z$
production, $\ee \ra Zq\bar{q}$, is small and can be further reduced by
flavor tagging. In the high mass range where the decay $H \ra WW^*$ is
dominant, the main background is triple gauge boson production and is
suppressed by two powers of the electroweak coupling. 

The $WW$ fusion mechanism, $\ee \ra \nu \bar{\nu}H$, offers a complementary 
production channel. For small $M_H$, the main backgrounds are single $W$ 
production, $\ee \ra e^\pm \nu W^\mp$ $[W \ra q\bar{q}$ and the $e^\pm$ 
escape detection] and $WW$ fusion into a $Z$ boson, $\ee \ra \nu 
\bar{\nu}Z$, which have cross sections 60 and 3 times larger than the
signal, respectively. Cuts on the rapidity spread, the energy and
momentum distribution of the two jets in the final state [as well as 
flavor tagging for small $M_H$] will suppress these background events. 

It has been shown in detailed simulations, that just a few fb$^{-1}$ 
of integrated luminosity are needed to obtain a 5$\sigma$ signal for
a Higgs boson with a mass $M_H \sim 140$ GeV at a 500 GeV collider 
[in fact, in this case, it is better to go to lower energies where the 
cross section is larger], even if it decays invisibly [as it could 
happen in SUSY models for instance]; Fig.5a. Higgs bosons with masses 
up to $M_H \sim 350$ GeV can be discovered at the 5$\sigma$ level, in both 
the strahlung and fusion processes at an energy of 500 GeV and with a
luminosity of 50 fb$^{-1}$; Fig.5b. For even higher masses, one needs to 
increase the c.m. energy of the collider, and as a rule of thumb, Higgs 
masses up to $\sim 70\%$ of the total energy of the collider can be 
probed. This means than a $\sim 1$ TeV collider will be needed to probe the 
entire Higgs mass range in the SM. 

\vspace*{3mm}
\nn {\bf c) Determination of Higgs Properties}
\vspace*{3mm}

Once the Higgs boson is found it will be of great importance to explore 
all its fundamental properties. This can be done at great details in
the clean environment of $\ee$ linear colliders: the Higgs mass, the
spin and parity quantum numbers and the couplings to fermions and
gauge bosons can measured, as discussed below.

$\bullet$ In the Higgs--strahlung process with the $Z$ decaying into
visible particles, the mass resolution achieved with kinematical
constraints is close to 5 GeV, and a precision of about $\pm 200$ MeV
can be obtained on the Higgs mass with $\int {\cal L}=10$
fb$^{-1}$ if the effects of beamstrahlung are small. 
For masses below 250 GeV, the Higgs boson is extremely narrow and its
width cannot be resolved experimentally; only for higher masses [or 
possibly in extended models], $\Gamma_H$ can be measured. 

$\bullet$ The angular distribution of the $Z/H$ in the Higgs--strahlung 
process is sensitive to the spin--zero of the Higgs particle: 
at high--energies the $Z$ is
longitudinally polarized and the distribution follows the $\sim
\sin^2\theta$ law which unambiguously characterizes the production of a
$J^P=0^+$ particle. The spin--parity quantum numbers of the Higgs
bosons can also be checked experimentally by looking at correlations in
the production $\ee \ra HZ \ra$ 4--fermions or decay $H \ra WW^* \ra$
4--fermion processes, as well as in the more difficult
channel $H \ra \tau^+ \tau^-$ for $M_H \lsim 140$ GeV. An unambiguous test
of the CP nature of the Higgs bosons can be made in the process $\ee \ra
tt \bar{H}$ or at laser photon colliders in the loop 
induced process $\gamma \gamma \ra H$.

$\bullet$ The masses of the fermions are generated through the Higgs
mechanism and the Higgs couplings to these particles are proportional to
their masses. This fundamental prediction has to be verified
experimentally. The Higgs couplings to $ZZ/WW$ bosons can be
directly determined by measuring the production cross sections in the
bremsstrahlung and the fusion processes. In the $\ee \ra H\mu^+\mu^-$
process, the total cross section can be measured with a precision of
less than 10\% with 50 fb$^{-1}$. 

The Higgs couplings to light fermions are harder to measure,
except if $M_H \lsim 140$ GeV. The Higgs branching ratios to $b\bar{b}$,
$\tau^+\tau^-$ and $c\bar{c}+gg$ can be measured with a precision of
$\sim 5, 10$ and $40 \%$ respectively for $M_H \sim 110$ GeV. 
For $M_H \sim 140$ GeV, BR$(H \ra WW^*)$ becomes sizeable and can be
experimentally determined; in this case the absolute magnitude of the
$b$ coupling can be derived since the $HWW$ coupling is fixed by the
production cross section. 
The Higgs coupling to top quarks, which is the largest coupling in
the electroweak theory, is directly accessible in the process $\ee \ra
t\bar{t}H$. For $M_H \lsim 130$ GeV, $\lambda_t$ can be
measured with a precision of about 10 to 20\% at $\sqrt{s}\sim 500$ GeV
with $\int {\cal L} \sim 50$ fb$^{-1}$. For $M_H \gsim 350$ GeV, the
$Ht \bar{t}$ coupling can be derived by measuring the $H \ra t\bar{t}$
branching ratio at higher energies. 

Finally, the measurement of the trilinear Higgs self--coupling,
which is the first non--trivial test of the Higgs potential, is
possible in the double Higgs production processes $\ee \ra ZHH$ and
$\ee \ra \nu \bar{\nu}HH$. However, the cross sections are rather
small and very high luminosities [and very high energies in the second
process] are needed. 

\section{Extensions of the SM}

\subsection{Supersymmetry}

An even stronger case for $\ee$ colliders in the 300--500 GeV energy
range is made by supersymmetric theories. The minimal supersymmetric
extension of the Standard Model (MSSM) requires the existence of two
isodoublets of Higgs fields, leading to three neutral, $h/H$(CP=+),
$A$(CP=--) and a pair of charged scalar particles. Besides the four
masses, two additional parameters define the properties of these
particles: a mixing angle $\alpha$ in the neutral CP--even sector and
the ratio of the two vacuum expectation values $\tb$, which from GUT
restrictions is assumed in the range $1 < \tb <m_t/m_b$. Supersymmetry
leads to several relations among these parameters and only two of them
[in general $\tb$ and $M_A$] are in fact independent. In the MSSM, the
upper bound on the mass of the lightest Higgs boson $h$ is shifted from
the tree level value $M_Z$ to $\sim 140$ GeV. The
masses of the heavy neutral and charged Higgs particles can be expected,
with a high probability, in the range of the electroweak symmetry
breaking scale. Some of these features are not specific to the minimal
extension and are expected to be realized also in more general SUSY
models. For instance, a light Higgs boson with a mass below ${\cal
O}$(200 GeV) is quite generally predicted by SUSY theories. 

At hadron colliders, the search for the Higgs bosons of the MSSM will be
even more difficult than for the SM Higgs boson. This is mainly due to
the fact that the important decays of the neutral Higgs particles into
$\gamma \gamma$ and $ZZ$ final states have in general smaller branching 
ratios,
especially if decays into SUSY particles are kinematically allowed. If
$M_h$ is close to its maximum value, $h$ has SM like couplings and the
situation is similar to the SM case with $M_H \sim $ 100--140 GeV. 

In $\ee$ collisions, besides the usual Higgs--strahlung and fusion
processes for the production of the CP--even Higgs bosons $h$ and $H$,
the neutral Higgs particles can also be produced pairwise: $\ee \ra A +
h/H$. The cross sections [Fig.~6a] for the Higgs--strahlung and the pair
production as well as the cross sections for the production of $h$ and
$H$ are mutually complementary, coming either with a coefficient
$\sin^2(\beta- \alpha)$ or $\cos^2(\beta -\alpha)$. The cross section
for $hZ$ production is large for large values of $M_h$, being of ${\cal
O}(50$ fb); by contrast, the cross section for $HZ$ is large for light
$h$ [implying small $M_H$].  In major parts of the parameter space, the
signals consist of a $Z$ boson and a $b\bar{b}$ or a $\tau^+ \tau^-$
pair, which is easy to separate from the main background, $\ee \ra ZZ$
[for $M_h \simeq M_Z$, efficient $b$ detection is needed]. For the
associated production, the situation is opposite: the cross section for
$Ah$ is large for light $h$ whereas $AH$ production is preferred in the
complementary region.  The signals consist mostly of four $b$ quarks in
the final state, requiring efficient $b\bar{b}$ quark tagging; mass
constraints help to eliminate the QCD jets and $ZZ$ backgrounds. The
CP--even Higgs particles can also be searched for in the $WW$ and $ZZ$
fusion mechanisms. Charged Higgs bosons can be produced pairwise, $\ee
\ra H^+H^-$ through $\gamma,Z$ exchange. The cross section depends only
on $M_{H^\pm}$ and is large up to $M_{H^\pm} \sim 230$~GeV.
Charged Higgs bosons can also be produced in top decays as discussed
previously. 

The discussion on the MSSM Higgs sector in $\ee$ linear colliders can be
summarized in the following points: i) The Higgs boson $h$ can be
detected in the {\it entire} range of the MSSM parameter space, either through
the bremsstrahlung process or through pair production; Fig.~6b. In fact,
this conclusion holds true even at a c.m. energy of 300 GeV and with a
luminosity of a few fb$^{-1}$. ii) There is a substantial area of the ($
M_h,\tb$) parameter space where {\it all} SUSY Higgs bosons can be
discovered at a 500 GeV collider; Fig.~6b. This is possible if the $H,A$ and
$H^{\pm}$ masses are less than $\sim 230$ GeV. For Higher masses, one
simply has to increase the c.m. energy. iii) In some parts of the MSSM
parameter space, the lightest Higgs $h$ can be detected, but it cannot
be distinguished from the SM Higgs boson. In this case, Higgs production
in $\gamma \gamma$ fusion [which receives extra contributions from SUSY
loops] can be helpful. iv) The properties of the Higgs bosons can
also be accurately determined; for instance, in the case of $h$, the same
tests as for the SM Higgs can be made. 

The lightest neutralinos and charginos, which are mixtures of the
supersymmetric partners of the electroweak gauge bosons and Higgs
bosons, are expected to be the lightest supersymmetric particles
[especially the lightest neutralino, which in the MSSM with conserved
R--parity, is the lightest SUSY particle and is stable]. To avoid
unnatural fine tuning in radiative corrections, these particles 
should have masses below the Fermi scale. 

Neutralinos and charginos are difficult to find at hadron colliders, but
they are easy to detect at $\ee$ colliders. They are produced in pairs,
$\ee \ra \chi_i^+ \chi_j^-$ ($i,j=1,2)$ and $\ee \ra \chi_i^0 \chi_j^0$
$(i,j=1$--4), through $s$--channel $\gamma, Z$ exchange, and
$t$--channel sneutrino or selectron exchange. The cross sections can be
rather large, ${\cal O}(100$ fb) at $\sqrt{s}\sim 500$ GeV, and enough
events will be produced to discover these particles and study their
properties. Detailed experimental simulations have shown that these
particles can be found with masses up to the beam energy, if the mass
difference with the lightest neutralino is not less than 20 GeV, 
substantialy improving on the discovery reach of LEP2. 

Left or right--handed scalar particles correspond to each chiral SM 
fermion. Starting with a universal scalar mass $m_0$ at the GUT scale,
squark and slepton masses evolve differently down to low energies, the 
later being significantly smaller than the former [with the possible
exception of the stop squark]. The SUSY partners of leptons may have masses
below the Fermi scale. In $\ee$ collisions, sleptons are produced 
pairwise $\ee \ra \tilde{l} \tilde{l}$, through $s$--channel $\gamma,
Z$ exchange. For sneutrinos only $Z$ exchange is present, for selectrons
and electronic sneutrinos additional chargino/neutralino $t$--channel
exchanges are present. The cross sections are large, $\sigma \gsim 50$
fb for $\sqrt{s} \sim 500$ GeV, and these states can be discovered
up to the kinematical limit if the slepton masses are larger than the 
lightest neutralino mass by more than a few ten GeV.

While colored squarks [except for the stop squarks] and gluinos can be 
detected up to masses of ${\cal O}(1$ TeV) at LHC, the sleptons 
are very hard to detect due to the low production rates and the 
large backgrounds, so that $\ee$ colliders are unique in this sector. 
In $\ee$ collisions, one can also measure the masses of the scalar
particles [which is difficult at hadron colliders because of the
escaping LSP's] and their couplings. This is very important in
order to constrain the various SUSY scenarios. 

Finally, due to the large value of $m_t$, the lightest stop quark 
can be lighter than the top quark and all the other squarks. Because of
the large backgrounds, stop squarks are also difficult to find at the 
LHC while they can be easily detected in $\ee$ collisions with masses 
up to the beam energy. 

\subsection{Grand Unified Theories} 

The SM does not unify the electroweak and strong forces in a
satisfactory way since the couplings of these interactions are
different. Therefore one would expect that a more fundamental theory
exists which describes the three forces within the context of a single
gauge group [which will contain the SM as a subgroup and will reduce to
it at low energies] and hence, with only one coupling constant.
The LEP data show that this can be indeed achieved in
Supersymmetric Grand Unified Theories. 

Two predictions of GUT can have dramatic phenomenological consequences in 
the ${\cal O}$(TeV) energy range: (i) The unifying group must be 
spontaneously broken at the unification scale for the proton to be
stable; however, it is possible that the breaking to the SM group 
occurs in several steps and that some subgroups remain unbroken down to 
a scale of order 1 TeV allowing for new neutral gauge bosons $(Z')$ 
with masses not far 
from the Fermi scale. (ii) The GUT groups incorporate fermion
representations	in which a complete generation of SM quarks and
leptons can be naturally embedded and in most cases these
representations are large enough 
to accomodate additional new fermions. These fermions, which are needed to 
have anomaly--free theories, can have masses not much larger than the
Fermi scale. In addition, particles with exotic quantum numbers such as
difermions [leptoquarks, diquarks and dileptons], could occur.

The direct search for these new matter and gauge particles and tests of
their indirect effects will be a major goal of the next generation of
accelerators and high--energy $\ee$ linear colliders have a rather rich 
potential for these searches.

If the energy can be raised high enough, the $\ee$ collider will operate
as a $Z$' factory; the event rates will be very high and the properties
of the $Z'$ can be studied in great details. A heavy $Z'$ boson, even if
its mass is substantially larger than the available center of mass
energy, will manifest itself through its propagator effects in the
process $\ee \ra \gamma, Z, Z' \ra$ fermions, producing potentially
sizeable effects on the leptonic cross section $\sigma^{\rm lept}$, the
ratio $R= \sigma^{\rm
had}/\sigma^{\rm lept}$, the forward--backward asymmetry $A_{\rm FB}^{
\rm lept}$ and if longitudinal polarization is available, the
left--right asymmetries $A_{\rm LR}^{\rm lept}$ and $A_{\rm LR}^{\rm
had}$. Masses up to 6 times the c.m. energy of the collider can be
probed for the expected luminosities. If a $Z'$ with mass below 3 TeV is
discovered at LHC, even a 500 GeV $\ee$ collider would give valuable
contributions to its detailed investigation by allowing the distinction
between different classes of models and the determination of the model
parameters. The two types of colliders would then provide complementary
information. 

$\ee$ colliders are well suited machines for the search of new leptons.
These particles can be pair produced, $\ee \ra L\bar{L}$, with large rates if
their masses are smaller than the beam energy. They can also be singly
produced in association with their standard light partners, $\ee \ra
L\bar{l}+\bar{L}l$ if the mixing between the light and heavy states is
not prohibitively small; one can then reach masses close to the total
energy of the collider. The signatures, with final states involving
charged leptons and gauge bosons, have clear characteristics so that
the detection of these particles should not be difficult in the clean
environment of $\ee$ colliders. Since they are strongly interacting
particles, quarks, leptoquarks and diquarks will be produced at the LHC
with very large rates. However, because of the difficult hadronic jet
background, the signals would be hard to analyze in detail. $\ee$
colliders would provide the ideal framework for highly precise analyses
of the properties of these new exotic particles if they are found at the
hadron colliders. 

\section{Conclusions}

$\ee$ linear colliders operating in the energy range of $\sim 500$ GeV
[and which can be extended to energies up to 2 TeV] have a very rich
physics potential, which in many aspects is complementary to that of
the LHC. High--precision measurements in the top quark and $W/Z$ bosons 
sectors can be performed, allowing further tests of the Standard Model and
eventually opening a window to new physics. The exploration of the
electroweak symmetry breaking mechanism can be made at great details:
Higgs particles can be easily searched for, and the clean environment of
the collider allows detailed studies of their basic properties. The
$\ee$ colliders provide a unique opportunity to explore in a deep 
manner central aspects of supersymmetric theories: the Higgs spectrum 
and its properties, the higgsino/gaugino and slepton sectors. 
Finally, gauge extensions of the Standard Model can be explored by 
searching for new gauge bosons and new matter particles. 

\vspace*{.5cm}

\nn {\bf Acknowledgements}: 

\nn I would like to thank the organizers of these Rencontres, in
particular Giorgio  Bellettini and Mario Greco, for their invitation,
their support and for the nice and stimulating atmosphere of the
meeting. This work is supported by DFG (Bonn).

\end{document}